\documentclass[12pt,
reprint,
aip,
floatfix,
]{revtex4-2}

\usepackage{upgreek}
\usepackage{bm}
\usepackage{xr}
\usepackage{graphicx}
\usepackage{gensymb}
\usepackage{float} 
\usepackage{xfrac}
\usepackage{extdash}
\usepackage{epsfig}
\usepackage{isomath}
\usepackage{textcomp}
\usepackage{bm}
\usepackage[english]{babel}
\usepackage{todonotes}
\usepackage{color,soul}
\usepackage{times}
\usepackage{mathtools}
\usepackage[separate-uncertainty=true]{siunitx}

\DeclareFontFamily{U}{euc}{}
\DeclareFontShape{U}{euc}{m}{n}{<-6>eurm5<6-8>eurm7<8->eurm10}{}%
\DeclareSymbolFont{AMSc}{U}{euc}{m}{n} 
\DeclareMathSymbol{\umu}{\mathord}{AMSc}{"16} 
\renewcommand{\vec}[1]{\boldsymbol{#1}}

\newcommand{\ensuretext}[1]{\ensuremath{\text{#1}}}

\newcommand{\textmutext}{\ensuretext{\textmu}}

\newcommand{\Mum}{\ensuremath{\umu}\ensuremath{\mathrm{m}}}
\newcommand{\mum}{\textrm{\,\ensuremath{\mathrm{\Mum}}}}
\mathchardef\ordinarycolon\mathcode`\:
\mathcode`\:=\string"8000
\begingroup \catcode`\:=\active
  \gdef:{\mathrel{\mathop\ordinarycolon}}
\endgroup

\begin{document}

\title{Demonstration of full-scale spatio-temporal diagnostics of solid-density plasmas driven by an ultra-short relativistic laser pulse using an X-ray free-electron laser}

\author{Lingen Huang$^{1}$}
\email{lingen.huang@hzdr.de}
\author{Michal Šmíd$^{1}$}
\author{Long Yang$^{1,2}$}
\author{Oliver Humphries$^{3}$}
\author{Johannes Hagemann$^{4}$}
\author{Thea Engler$^{4}$}
\author{Xiayun Pan$^{1,2}$}
\author{Yangzhe Cui$^{1}$}
\author{Thomas Kluge$^{1}$}
\author{Ritz Aguilar$^{1}$}
\author{Carsten Baehtz$^{1}$}
\author{Erik Brambrink$^{3}$}
\author{Engin Eren$^{5}$}
\author{Katerina Falk$^{1}$}
\author{Alejandro Laso Garcia$^{1}$}
\author{Sebastian G\"{o}de$^{3}$}
\author{Christian Gutt$^{6}$}
\author{Mohamed Hassan$^{1}$}
\author{Philipp Heuser$^{5}$}
\author{Hauke H\"{o}ppner$^{1}$}
\author{Michaela Kozlova$^{1}$}
\author{Wei Lu$^{3}$}
\author{Josefine Metzkes-Ng$^{1}$}
\author{Masruri Masruri$^{1}$}
\author{Mikhail Mishchenko$^{3}$}
\author{Motoaki Nakatsutsumi $^{3}$}
\author{Masato Ota $^{7}$}
\author{\"{O}zg\"{u}l \"{O}zt\"{u}rk$^{6}$}
\author{Alexander Pelka$^{1}$}
\author{Irene Prencipe$^{1}$}
\author{Thomas R. Preston$^{3}$}
\author{Lisa Randolph$^{3}$}
\author{Martin Rehwald$^{1}$}
\author{Hans-Peter Schlenvoigt$^{1}$}
\author{Ulrich Schramm$^{1,2}$}
\author{Jan-Patrick Schwinkendorf$^{3}$}
\author{Sebastian Starke$^{1}$}
\author{Radka Štefaníková$^{1,2}$}
\author{Erik Thiessenhusen$^{1}$}
\author{Monica Toncian$^{1}$}
\author{Toma Toncian$^{1}$}
\author{Jan Vorberger$^{1}$}
\author{Ulf Zastrau$^{3}$}
\author{Karl Zeil$^{1}$}
\author{Thomas E. Cowan$^{1,2}$}

\affiliation{%
1. Helmholtz-Zentrum Dresden-Rossendorf, 01328 Dresden, Germany\\
2. Technische Universit\"at Dresden, 01062 Dresden, Germany\\
3. European XFEL, 22869 Schenefeld, Germany\\
4. Center for X-ray and Nano Science CXNS, Deutsches Elektronen-Synchrotron DESY, Notkestraße 85, 22607 Hamburg, Germany \\
5. Helmholtz Imaging, Deutsches Elektronen-Synchrotron DESY, 22607 Hamburg, Germany\\
6. Universit\"{a}t Siegen, 57072 Siegen, Germany\\
7. National Institute for Fusion Sciences, 509-5292, Toki, Japan
}%


\date{\today}




\begin{abstract}

Understanding the complex plasma dynamics in ultra-intense relativistic laser-solid interactions is of fundamental importance to the applications of laser plasma-based particle accelerators, creation of high energy-density matter, understanding of planetary science and laser-driven fusion energy. However, experimental efforts in this regime have been limited by the accessibility of over-critical density and spatio-temporal resolution of conventional diagnostics. Over the last decade, the advent of femtosecond brilliant hard X-ray free electron lasers (XFELs) is opening new horizons to break these limitations. Here, for the first time we present full-scale spatio-temporal measurements of solid-density plasma dynamics, including preplasma generation with tens of nanometer-scale length driven by the leading edge of a relativistic laser pulse, ultrafast heating and ionization at the main pulse arrival, laser-driven blast shock waves and transient surface return current-induced compression dynamics up to hundreds of picoseconds after interaction. These observations are enabled by utilizing a novel combination of advanced X-ray diagnostics such as small-angle X-ray scattering (SAXS), resonant X-ray emission spectroscopy (RXES), and propagation-based X-ray phase-contrast imaging (XPCI) simultaneously at the European XFEL-HED beamline station. 

\end{abstract}
\maketitle 

\section{Introduction}

When irradiating a solid target by an ultra-short relativistic laser pulse with peak intensity in the order of 10$^{20}$ W/cm$^{2}$, a copious number of electrons within a few nanometers (nm) skin depth are promptly ionized and accelerated up to few tens of MeV kinetic energy. The relativistic electrons then propagate ballistically into the solid target with a large current density in the order of 10$^{13}$ A/cm$^{2}$, depending on the laser absorption\cite{Huang2016}. It is well-known that the laser absorption during the arrival of the relativistic laser peak is critically determined by the surface density profile of the preplasma caused by the intrinsic laser rising edge or the prepulse, a few tens of picoseconds (ps) before the peak of the pulse is reached\cite{Bagnoud2017,Bernert2023}. Although numerous studies have been conducted to measure the absolute laser absorption efficiency and its correlation to the scaling of generated hot electron temperature\cite{Wilks1992,Beg1997,Klug2011,Park2021}, it is still challenging to directly probe the density gradient of preplasma due to its small spatial scale. The transport of such an intense hot electron beam through the solid target can further trigger abundant ultrafast plasma dynamics such as isochoric heating and ionization, development of kinetic instabilities, generation of extreme fields, radiation sources and warm/hot dense plasmas. Moreover, after the main laser pulse reaches the target surface, hydrodynamic processes of shock generation, heat diffusion, magnetic Z-pinch and thermal pressure driven compression, magneto-hydrodynamic instabilities occur over few-ps to sub-ns timescales. Towards understanding the underlying physics of electron transport, a tremendous effort has been put to diagnose these complex phenomena via a variety of diagnostics including optical shadowgraphy\cite{Bernert2022}, charged particle imaging\cite{Quinn2009,Quinn2012}, X-ray radiography\cite{Morace2014}, X-ray self-emission spectroscopy\cite{Zastrau2010} and X-ray Thomson scattering (XRTS)\cite{Glenzer2009} over last several decades. However, these diagnostics are limited either by inaccessibility to solid density, harsh Bremsstrahlung background noise, or insufficient temporal and small spatial scales. Therefore, numerical approaches which can be more readily interrogated and performed at scale such as magneto-hydrodynamics (MHD) and particle-in-cell (PIC) simulations play a key role of fundamental understanding of the complex plasma dynamics\cite{Huang2017,Yang2024}. Intuitively, to precisely manipulate and control the high-intensity laser-solid interactions for the potential applications including cancer therapy via compact plasma-based particle accelerator\cite{Kroll2022} and laser-driven fusion energy\cite{NUCKOLLS1972,Kodama2001}, it is highly desirable to develop novel diagnostics to overcome the limitations, and thus benchmark and strengthen the predictive power of numerical simulations. 

\begin{figure*}
    \centering
    \includegraphics[width=\textwidth]{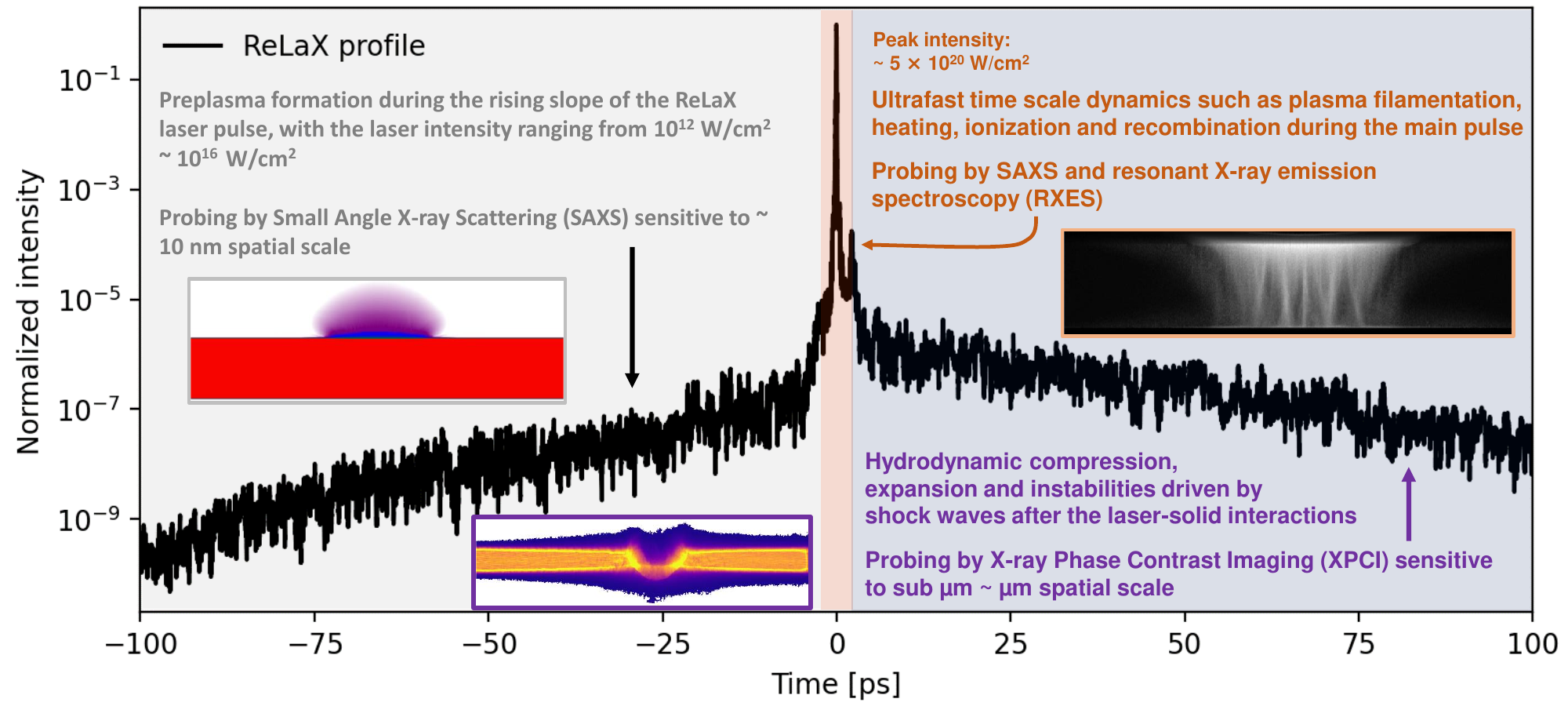}
    \caption{A schematic of the multi-scale spatio-temporal plasma processes induced by electron transport in solid-density plasmas irradiated by ultra-short, relativistic laser pulses, along with the corresponding X-ray diagnostics used in the pump-probe experiment. The black curve shows the realistic temporal profile of the ReLaX laser pulse ranging from -100 ps to 100 ps. The illustration includes the preplasma formation, electron filamentation, blast wave and compression, as predicted by our previous simulations\cite{Huang2017,Yang2024}, is depicted along with the curve and text.}
    \label{fig:fig1}
\end{figure*}

The latest advent of novel experimental platforms equipped with both optical high power laser and ultra-short brilliant X-ray free electron lasers (XFELs) such as European XFEL-HED\cite{Zastrau2021}, LCLS-MEC\cite{Nagler2015} and SACLA beamline\cite{Yabuuchi2019}, provides unprecedented opportunities for experimental studies of solid density plasmas. State of the art X-ray sources with an extremely high photon number per pulse (in the order of 10$^{12}$), ultra-short duration (tens of fs) and wavelength (0.5 \r{A}) exhibiting nearly full transverse coherence, enables probing the surface and inner structures of such overdense plasmas via scattering, diffraction and absorption imaging, as well as spectroscopic methodologies in femtosecond and nanometer resolution simultaneously utilizing a single shot. The recent pioneering work conducted at the current hard XFEL facilities combining femtosecond high power lasers using the X-ray diagnostics of small angle X-ray scattering (SAXS)\cite{Kluge2018,Gaus2020,Kluge2023}, grazing incidence SAXS (GISAXS)\cite{Randolph2022,Randolph2024}, X-ray phase contrast imaging (XPCI)\cite{Schropp2015,Hagemann2021,Hodge2022,Garcia2024}, absorption imaging\cite{Sawada2023,Sawada2024} and XRTS\cite{Frydrych2020} clearly demonstrate the excellent capabilities. 

Here, combining the cutting-edge diagnostics of SAXS, XPCI and resonant X-ray emission spectroscopy (RXES), for the first time we are able to probe the complex dynamics of electron transport in the relativistic laser-solid interactions in a multi-scale spatio-temporal regime ranging from few nanometers up to micron and few tens of femtoseconds to hundreds of picoseconds. The pump-probe experiment is performed at the European XFEL-HED station, using an optical laser with the peak intensity up to $5\times10^{20}$ W/cm$^{2}$ irradiating on 10 to 25 $\textmutext$m diameter Cu and other wires probed by a femtosecond brilliant hard XFEL. In the experiment, we observe a significant decrease of SAXS yield during the arrival of the leading edge of ReLaX's laser pulse from 40 ps to 1 ps (ReLaX refers to the high-intensity laser system known as the Relativistic Laser at EuXFEL,  with its temporal profile illustrated in figure 1). This clearly indicates the formation and expansion of a preplasma by the pulse pedestal with intensity rising from $10^{13}$ W/cm$^{2}$ to $10^{16}$ W/cm$^{2}$. While SAXS is sensitive to a few nm structure changes, making it appealing to probe the formation of preplasma and also development of electron filamentation during the pre- and intra-short-laser-pulse solid interactions\cite{Ordyna2024}, XPCI is sensitive to the density gradient with sub-$\textmutext$m to $\textmutext$m spatial scale, enabling it to probe hydrodynamics driven density evolution after laser-solid interactions such as the growth of hydrodynamic instabilities, thermal expansion and shock compression in tens of ps to sub-ns. The complementary X-ray diagnostics have a great potential to fully measure the spatio-temporal evolution of the plasma density. In our experiment, the measured XPCI patterns in the case of a 30 $\textmutext$m field of view (FoV) illumination the  Cu wire of 10 $\textmutext$m diameter exhibit asymmetric contrast fringes on either side of the wire due to the non-uniform distribution of the XFEL intensity. The diffraction pattern changes significantly till target destruction at $\sim$150 ps after the laser-wire interaction. The reconstructed phase map shows the feature of laser-driven blast shock waves and transient surface return current-induced expansion and compression, which has recently been measured with direct X-ray imaging\cite{Garcia2024}. However, precise reconstruction of the plasma density distributions is quite challenging since not all the constraints could be applied in the phase retrieval algorithm limited by the data quality and uncertainties of the propagation parameters. In conjunction with SAXS and XPCI measuring the complementary information of density evolution that are sensitive to few nm and $\mum$ scales respectively,  time-resolved resonant X-ray emission spectroscopy (RXES) is used to probe the dynamic plasma heating and ionization during the main pulse-wire interactions. The fluorescence X-ray emission spectroscopy (XES) has proven to be one of the most valuable diagnostics to determine the plasma temperature and ionization by analyzing the intensity and ratio of characteristic X-ray lines at specific energies through such as K$_{\alpha}$, K$_{\beta}$ and He$_{\alpha}$ emissions\cite{Beier2022}. But due to the extremely short femtosecond time scale evolution of the dense plasmas driven by the ultra-fast ultra-intense laser, the measured spectroscopic data are usually time integrated. Coupling an X-ray streak camera in the XES is able to measure the time-resolved emission spectroscopy, but currently it is limited to $\sim$1 ps temporal resolution\cite{Humphries2020}. The advent of ultra-short XFELs with femtosecond intense fields provide unprecedented temporal resolution to probe the transient non-equilibrium plasma states under extreme conditions driven by the high-intensity optical lasers,  which is limited only by the XFEL pulse duration and the timing stability between XFEL and optical lasers (fs to few tens of fs)\cite{Zastrau2021,Osterhoff2021}. In this experiment, the XFEL photon energy is set at 8.2 keV that is in resonance with the bound-bound transition of highly charged nitrogen-like Cu ions, \textit{i.e.}, Cu$^{22+}$(7 bound electrons in the inner shells), pumped by the optical relativistic laser. According to the atomic collision-radiative simulations, the XFEL radiation field is intense enough to redistribute the electronic configuration in the isoelectronic sequences (same Cu nuclear charge but with different electronic configurations) when the specific charged state Cu$^{22+}$ is present in the hot dense plasmas. Then the inner K-shell electrons are resonantly excited to the L-shell vacancies by absorbing the XFEL energy, followed by the processes of recombination, spontaneous and stimulated X-ray emission. The experimentally observed enhancement in XFEL beam attenuation and resonant X-ray emission provides a direct indication of the population of the selected ionization state. Accordingly, the measurement of spectroscopic evolution from hot K$_{\alpha}$ emission to \textit{K–L} resonant absorption reflects the underlying heating, ionization, and recombination dynamics, and can serve as a valuable benchmark for validating collisional and ionization models in solid-density plasmas\cite{Kluge2016,Huang2017,Gaus2020}.

In brief, combing multiple advanced X-ray diagnostics of SAXS and PCI, in conjunction with RXES, we are able to measure the preplasma formation and expansion, laser driven blast wave and ablation driven cylindrical compression launched by the surface transient current sensitive to few nm-$\textmutext$m scales, as well as the transient heating and ionization in tens of fs temporal resolution simultaneously. It enables us to fully understand the multi-spatio-temporal dynamics of electron transport through the solid density plasmas irradiated by the ultra-short relativistic laser pulses, as illustrated in figure 1 together with the realistic temporal profile of the ReLaX laser pulse ranging from -100 ps to 100 ps. 

\section{Experimental setup}

The pump-probe experiment was performed at the HED endstation located at the European XFEL using the ultra-intense optical laser facility ReLaX to create the solid density plasmas, that is a 3 J, 30 fs, 10 Hz Ti:sapphire based system\cite{Laso2021}. The schematic illustration of the experimental setup is shown in figure 2. The ReLaX laser is focused to a FWHM spot size of approximately 4 $\mum$ measured by the In Line Microscope (ILM) diagnostics with 20$\times$ microscope objective, resulting the peak intensity up to $5\times10^{20}$ W/cm$^{2}$ irradiating on the wire samples by varying the materials (Cu, plastic coated Cu and tungsten) and diameters from 5 to 35 $\mum$ at normal incidence. The temporal contrast of pulse intensity shown in figure 1 is measured by the offline diagnostics package consisting of single-shot temporal pulse diagnostics, a scanning third-order intensity autocorrelator and a spatial phase sensor in combination with full beam adaptive deformable mirror\cite{Laso2021}. 

The XFEL beam generated by the principle of self-amplified spontaneous emission (SASE) with the energy of $\sim$1.5 mJ, FWHM pulse duration of $\sim$25 fs and photon energy centered at 8.2 keV with a FWHM bandwidth of $\sim$20 eV is used to probe the solid density plasmas, aided by the combination of multiple X-ray diagnostics including both imaging and spectroscopy such as SAXS, RXES and XPCI simultaneously. The online calibrated HIgh REsolution hard X-ray single-shot (HIREX) spectrometer installed in the photon tunnel of the SASE2 undulator beamline was used to monitor the SASE spectrum\cite{Kujala2020} that ensures the central X-ray photon energy fixed at 8.2 keV during the entire beamtime. The photon energy was cross checked by the elastic X-ray scattering signal on Cu samples measured by the von Hámos X-ray spectrometer inside the interaction chamber\cite{Preston2020}. The X-ray pulse energy is measured via an X-ray gas monitor (XGM) located in the photon tunnel and an intensity and position monitor (IPM) installed at the HED instrument\cite{Zastrau2021}. The energy calibration of IPM is performed during the preparation for the hot shots each beamtime shift. 

Two sets of X-ray Compound Refractive Lens (CRL) configurations are used to achieve small and large field of view (FoV) on the sample that are preferred for the SAXS/RXES and XPCI diagnostics respectively. In the case of small FoV, arm 3, 6 and 10 of CRL3 (located in the HED optics hutch at $\sim$962 m from the source\cite{Zastrau2021}) are set to get FWHM spot size of $\sim10 \mum$ on the sample. In large FoV, arm 4 and 6 of CRL3  are used to achieve FWHM spot size of $\sim 30 \mum$ on the sample. The XFEL focal spot size is characterized by the edge scan of test samples via X-ray transmission imaging, using a GAGG scintillator-based detector (Optique Peter) coupled to an Andor Zyla CMOS camera equipped with a 7.5× objective lens, positioned 6.31 meters downstream from the target chamber center. Two Highly Annealed Pyrolytic Graphite(HAPG) mirrors located symmetrically below and above the propagation path of XFEL beam are used to reflect the SAXS signal to the Jungfrau X-ray detector\cite{Jungfrau}. Simultaneously, the XFEL beam passing through the gap between the HAPG mirrors is recorded by Optique Peter for XPCI imaging\cite{Smid2020}.

The timing and synchronization between optical and X-ray laser was measured by the HED optical en-coding spatial photon arrival monitor (PAM)\cite{Kirkwood2019}. The measurable timing window by the PAM is within 200 fs, giving upper limit of the uncertainty of XFEL probe time delay relative to the ReLaX laser presented in this work. 

\begin{figure*}
    \centering
    \includegraphics[width=\textwidth]{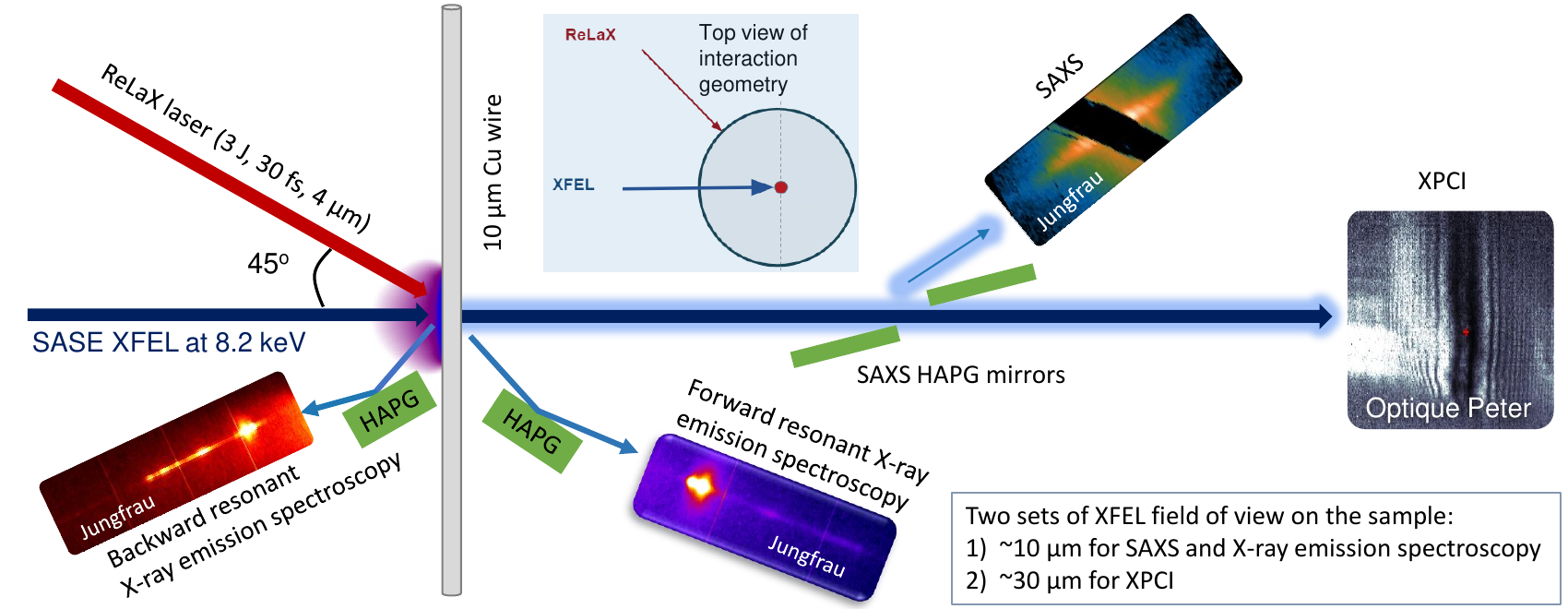}
    \caption{A schematic illustration of the experimental setup. Both optical relativistic ReLaX laser and XFEL irradiate the Cu wire of 10 $\mum$ diameter at normal incidence, polarized in the horizontal direction. The beams intersect at an angle of  the ReLaX and X-ray beam is \ang{45}. The SAXS patterns and emission spectroscopy are recorded by the Jungfrau X-ray detectors. Two HAPG chromatic mirror with a small gap are used to reflect the SAXS signal, allowing the propagation of primary XFEL beam to an Andor Zyla CMOS camera via a 7.5$\times$ objective for simultaneous XPCI imaging. The X-ray propagation distances from the sample to the SAXS and XPCI detectors are 1.31 m and 6.31 m respectively. The entire setup except the SAXS Jungfrau detector and the last 50 cm before the Zyla detector is placed in vacuum conditions to minimize the air scattering. The SAXS/RXES and XPCI data presented in this study correspond to the small FoV and large FoV on the wires of $\sim$10 $\mum$ and $\sim$30 $\mum$ respectively, which are achieved by switching two sets of compound refractive lens (CRL) configurations as described in the main text.}
    \label{fig:fig2}
\end{figure*}

\section{Experimental Results and Discussions}

In the following sections we present the details of each plasma diagnostic in turn: nanometer scale-sensitive SAXS, charge state-sensitive RXES, and sub-$\mum$ scale-sensitive XPCI to fully investigate the integrated spatio-temporal dynamics of solid density plasmas driven by the ultra-intense laser pulse ReLaX. These diagnostic probe a range of spatio-temporal scales, including the preplasma generation with tens of nanometer-scale length driven by the leading edge of ReLaX laser pulse, ultra-fast heating and ionization at the main pulse arrival, direct laser driven blast shock waves, and transient surface return current induced thermal ablative shock driven cylindrical pinch dynamics up to hundreds of picoseconds after interaction. While various wire materials such as Cu, plastic coated Cu and tungsten with different diameters ranging rom 5 to 35 $\mum$ are shot in the experiment, only the case of 10 $\mum$ Cu wires is shown in this demonstration paper. The full data set will be presented in the follow-up publications.

\subsection{SAXS measuring the preplasma expansion}

\begin{figure*}
    \centering
    \includegraphics[width=\textwidth]{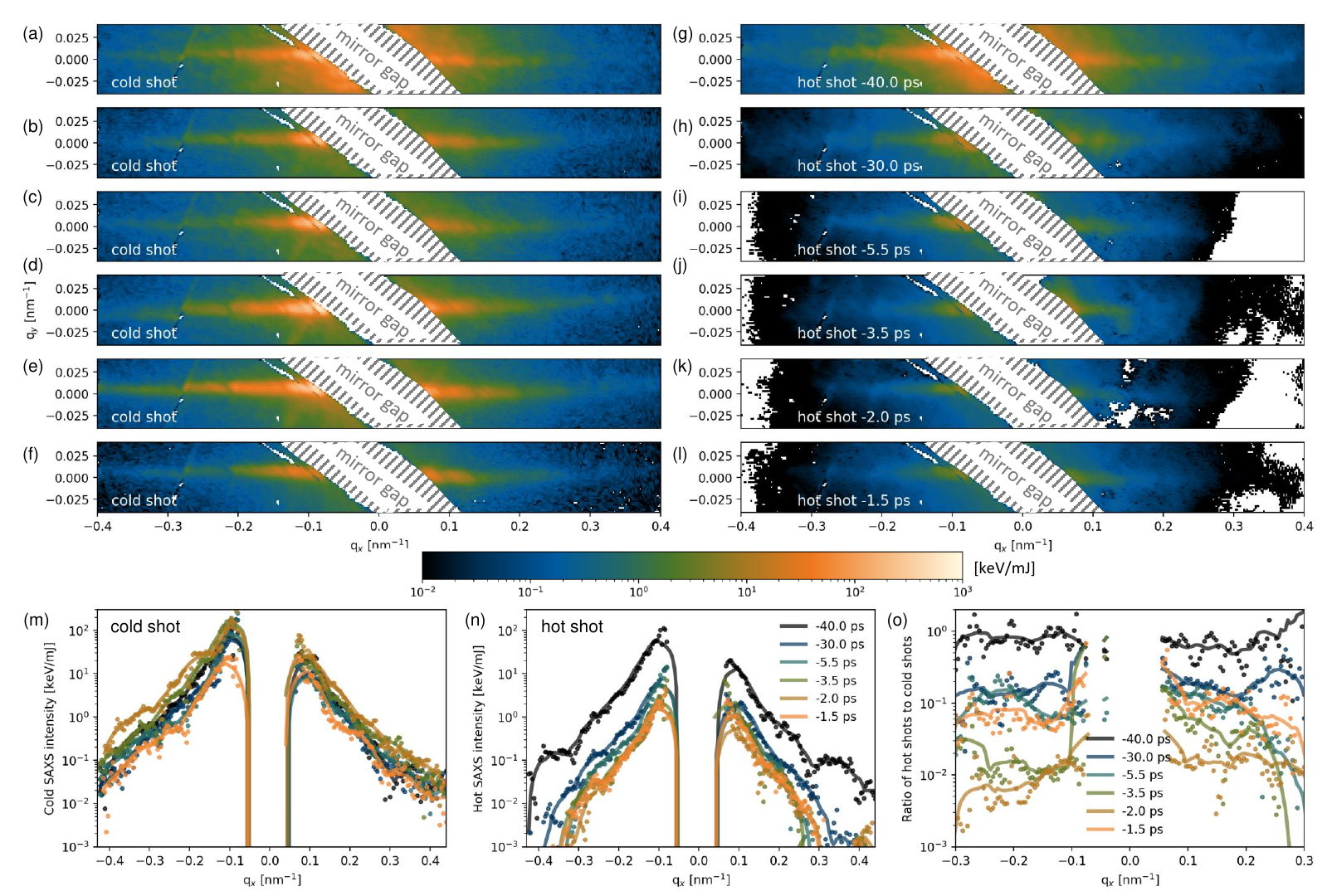}
    \caption{A summary of time resolved SAXS patterns (a-l) and the horizontal outlines taken through $q_y = 0$ averaging over 10 pixels (m-o) during the rising slope of the ReLaX laser from 40 ps to 1.5 ps prior to the arrival of the peak pulse intensity. The corresponding cold SAXS patterns are plotted as a reference of initial wire edge sharpness (a-f). The colorbar denotes the scattered X-ray energy normalized to the incoming XFEL energy.}
    \label{fig:fig3}
\end{figure*}

Small angle X-ray scattering arises from the coherent superposition of elastic X-ray scattering in a given direction - usually considered as $< 1^\circ$. This direction arises as a result of a momentum transfer $\vec{q}=\vec{k}-\vec{k}^{\prime}$, where $\vec{k}$ and $\vec{k}^{\prime}$ are the incoming and scattering wave vectors respectively. The SAXS intensity, $I(\vec{q})$, is proportional to the spatial Fourier transformation of the electron density, $n_e$, projected in the X-ray beam direction assuming homogeneous irradiation, as derived in previous theoretical work\cite{Nielsen2011,Kluge2014}:
\begin{equation}
I(q)=I_0 \lvert FT \left\{n_e \right\} \rvert^2, \label{eq:SAXS1}
\end{equation}

where $I_0$ is a constant factor depending on the initial incident X-ray fluence and the solid angle of the SAXS detector. Here we assume the projected density $n_e$ is the convolution of the form, structure and expansion terms that represent the averaged geometrical shape,  individual electron-electron correlations, and sharpness of the wire edge respectively, then the equation (1) can be re-written as: 
\begin{equation}
I(\vec{q})=I_0\cdot F(\vec{q})\cdot S\vec{q})\cdot E(\vec{q}), \label{eq:SAXS2}
\end{equation}

where $F(\vec{q})$, $S(\vec{q})$ and $E(\vec{q})$ are the form, structure and expansion factors in the reciprocal or momentum space respectively. In our SAXS setup, the momentum transfer $q$ ranges from 0.08 nm$^{-1}$ to 0.45 nm$^{-1}$, indicating that SAXS is able to measure the scale length of the correlation structure up from 5 nm to 100 nm. In this $q$ range of interest, the form factor of a wire is approximately equal to $F(q) \simeq 1/q^3$ according to Porod's law\cite{Porod1956,MelaniePhDThesis}. For a cold wire with infinitely sharp density gradient between the edge of the target and the surrounding vacuum (a Heaviside step function), the SAXS pattern exhibits a strong streak feature parallel to the surface normal direction, with the intensity proportional to the dominant form factor $I(q)\propto F(q)$. When the wire is irradiated by the rising slope ranging from a few tens of ps to 1 ps prior to the arrival of the peak ReLaX laser pulse, corresponding to the laser intensity rising from $\sim 10^{13}$ W/cm$^{2}$ to $10^{16}$ W/cm$^{2}$ as seen in figure 1, the scattering edge is smoothed by the preplasma expansion without bending by the hole boring process. Assuming the edge shape of wire radial density to be an error function $\mathrm{erf}\left(\frac{r}{\sqrt{2}\sigma}\right)$
with a finite scale length of $\sigma$ nm, the expansion factor then can be derived\cite{MasatoPhDThesis} as $E(q)=e^{-q^2\sigma^2/4}$. Since the laser intensity is far below the relativistic intensity, \textit{i.e.}, $10^{18}$ W/cm$^{2}$, the bulk electron density remains nearly constant because no energetic electrons with relativistic kinetic energy inject into the target to introduce any structure changes such as plasma oscillation and instabilities, thus the structure factor is close to unity, \textit{i.e.}, $S(q)\simeq 1$. Therefore, the function of SAXS intensity caused by the preplasma expansion can be simplified as: 
\begin{equation}
I(q)\simeq I_0\cdot \frac{1}{q^3}\cdot e^{-q^2\sigma^2/4}, \label{eq:SAXS3}
\end{equation}

Figure 3 shows the experimentally measured SAXS patterns and lineouts normalized by the incident XFEL energy from 40 ps to 1.5 ps prior to the arrival of the ReLaX peak laser intensity. The hot shots shown in figure 3 are selected based on similarly high He$_\alpha$ yield, indicating similar laser absorption efficiency for each shot due to the spatial jitter of the ReLaX laser spot, on the order of the focal spot size. Before each hot shot pumped by the ReLaX laser, we take a XFEL-only reference for the cold target geometry and XFEL-target overlap. 
Cold wire scattering seen to be consistent over the reference shots, both for the streaks in Fig.3(a)-(f), and the outlines in  Fig.3(m) over 4 orders of magnitude in signal; indicating uniform wire edge sharpness. Fitting the measured SAXS profile in Fig.3(m) with equation (3), yields a scale length or roughness below 10 nm. The intensity on the right SAXS streak is stable shot-to-shot. On the left streak, the intensity shows a relatively bigger variation. This asymmetric scattering feature is likely caused by the XFEL spatial jitter and randomly distributed nano-structure on the wire surface. For the hot shots, the 2D streak SAXS patterns shown in Fig.3(g)-(l) and the outlines in Fig.3(n) exhibit a clear trend of SAXS intensity decreasing with time delay. To take account of the initial density edge, Fig.3(o) shows the ratio of SAXS intensity of hot shot to cold shot that gives the same decreasing trend except for the time delay at -1.5 ps. It is surprising that the SAXS intensity decreases with one to two orders of magnitude from -40 ps to -1.5 ps within the entire measurable $q$ range. Such dramatic SAXS intensity reduction is not observed  in the case of sub-relativistic laser intensity \cite{Kluge2023}. It reveals that the preplasma is formed and expanded at least 30 ps (corresponding to the intensity of $\sim5\times10^{13}$ W/cm$^{2}$) prior to the arrival of main pulse with peak intensity of $5\times10^{20}$ W/cm$^{2}$. Besides the decreasing of SAXS intensity, we also observe that the slope of the right streak becomes slightly steeper which is correlated to the smoother edge caused by the pre-expansion. However, the slope of the left streak remains to be constant which makes it unreliable to extract the scale length of preplasma by fitting the SAXS data with equation 3. It infers that the dynamic density profile of the preplasma driven by the rising edge of the ReLaX laser is complex and can not be simply described by an error function. We notice that the experimental shot-to-shot uncertainties such as the fine structure of initial target surface, XFEL jitter, laser contrast fluctuations, overlapping of ReLaX laser, XFEL and wires can also influence the SAXS signal.

At the arrival of main laser pulse with relativistic intensity, the transport of fast electrons into the solid target can induce complex structure change by the processes of hole boring\cite{Kluge2023}, plasma heating and ionization\cite{Gaus2020}, oscillation and filamentation\cite{Ordyna2024} as inferred by our previous SAXS experiments. In this phase, the scattering features become much more complex that we have to take account into the contribution of structure factor $S(q)$. Additionally, the scattering intensity further diminishes to near the noise level, making it challenging to extract reliable dynamic information. Accessing SAXS in lower q region will improve the data quality, which is subject to our follow up experiment and analysis.

\subsection{RXES measuring heating and ionization dynamics}

During the interaction between the ultra-short laser and the solid Cu with preformed plasmas, a fraction of the electrons is accelerated to the relativistic kinetic energy and transport into the solid, leading to subsequent ultra-fast bulk heating, collisional and pressure ionization. The ionization then empties the outer electron shells of the atoms, and opens the possibility of resonant X-ray absorption in an optical laser pump-XFEL probe experiment, namely from the K shell (n=1) to a higher shell L (n=2) or M (n=3) transition, \textit{i.e.}, the inverse processes of $K_\alpha$ or $K_\beta$, as shown in the schematic figure 4(a). For a given transition, for example 1s to 2p, the binding energy depends on the number of remaining bound electrons. Due to the screening effect of the nuclear Coulomb potential, the higher charge state requires higher transition energy or resonant X-ray photon energy. The scenario is clearly seen from the opacity spectra of the optically thin solid density Cu plasmas simulated by the atomic collisional radiative code FLYCHK\cite{FLYCHK2005}, as shown in figure 4(b). In the calculated opacity spectrum, one sees a distribution of several satellite, slightly overlapping peaks. Each peak corresponds to a bound-bound transition for a specific charge state. It is well-known that the transition energy of cold $K_\alpha$ of Cu is 8.0478 keV according to the NIST data\cite{NIST2022}. As plasma temperature increases, the charge state becomes higher, thus the transition energy or the peak position of the spectrum shifts to higher photon energy. Based on this principle, an XFEL can be used to resonantly pump a specific atomic transition and thus get enhanced X-ray emission. The method has also been successfully applied to generate highly coherent atomic X-ray lasers pumped by an XFEL\cite{Rohringer2012,Yoneda2015}.  

In our experiment, we tune the photon energy of XFEL at 8.2 keV, that is mostly in resonance with the bound-bound transition energy of the selected specific charge state of Cu ions, \textit{i.e.}, Cu$^{22+}$. When the Cu is ionized to the certain charge state driven by the main laser pulse, the inner K shell electrons will be resonantly excited by the XFEL photon to the higher L shell electrons. These electrons then decay through the recombination process and re-emit the photons at the same energy as the incident X-rays exciting them in a very shot time, leading to enhanced X-ray emission yield at the resonant energy.  In case the solid-density plasma experiences continued heating, resulting in higher ionization degree than the selected charge state, the plasma's opacity diminishes due to the insufficiency of bound electrons to establish the resonant transition channel. The reduction of opacity also occurs when the plasma undergoes the cooling process, as shown in the temperature-dependent opacity of solid density Cu plasma at the specific photon energy $E_{photon} = 8.2$ keV in figure 4(c). 

\begin{figure*}
    \centering
    \includegraphics[width=\textwidth]{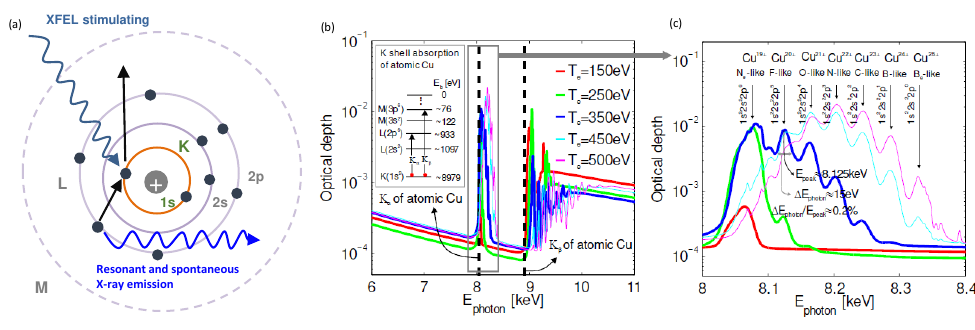}
    \caption{(a) Schematic of resonant X-ray emission stimulated by an external XFEL. (b) Optical depth of an optical transparent solid Cu plasma at different temperatures calculated by the atomic code FLYCHK over a broad photon energy and (c) over the energy region of interest in the solid density plasma\cite{HuangThesis}
    . Each satellite peak corresponds to a specific Cu ionic K-L bound-bound transition.}
    \label{fig:fig5}
\end{figure*}

\begin{figure*}
    \centering
    \includegraphics[width=\textwidth]{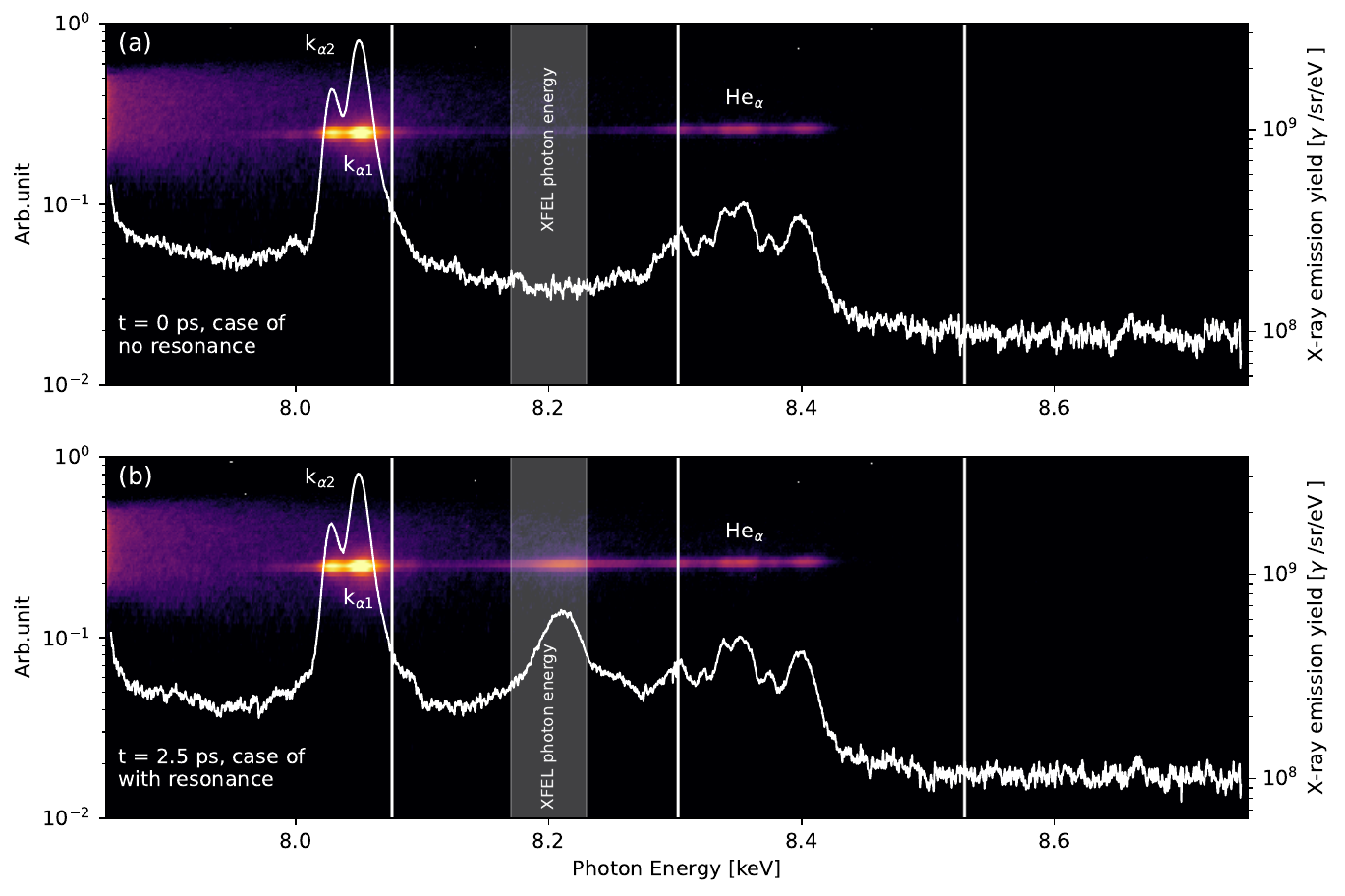}
    \caption{Experimentally measured time-resolved X-ray emission spectra in the case of no resonant yield at $\sim$ 0 ps (a) and the case of with strong resonant yield at $\sim$ 2.5 ps (b). The shaded gray region indicates the photon energy range of the SASE XFEL beam, centered at 8.2 keV with a narrow bandwidth of $\sim$40 eV, that is in resonance with the bound-bound transition of interest $1s^22s^22p^3$ to $1s^12s^22p^4$. The white vertical lines represent the gap between two adjacent chips of the Jungfrau detector .
    }
    \label{fig:fig6}
\end{figure*}

Figure 5 shows the measured X-ray emission spectra for two cases: one exhibiting no resonant yield at $\sim$ 0 ps, and the other showing a strong resonant yield at $\sim$2.5 ps. 
Prominent peaks corresponding to the cold K$_{\alpha}$ transitions (8.02–8.04 keV) and the  He$_{\alpha}$ transitions (8.28–8.42 keV) are observed in both cases, indicating comparable initial interaction conditions. As these transitions have no interference with the XFEL probe, their X-ray emission is time-integrated over the entire duration of the laser–solid interaction. Here, the number of measured K$_{\alpha}$ photons is $\sim(8.5\pm0.7)\times10^{10}$, that gives the total energy of $\sim1.36$ mJ assuming isotropic  K$_{\alpha}$ emission. This yields a conversion efficiency of $\sim4.2\times10^{-4}$ relative to the ReLaX laser energy. At $\sim$ 0 ps when the Cu is not highly ionized yet, the spectrum shows that the spontaneous X-ray fluorescence within the energy window covered by the XFEL bandwidth is weak due to the rapid heating. Its intensity is insufficient to surpass the harsh Bremsstrahlung environment background and thus buried in the continuum radiation. In the case stimulated by the intense XFEL at 2.5 ps, the population of the specific transition state is significantly enhanced due to the resonant absorption, leading to a distinct resonant peak well above the Bremsstrahlung background as shown in figure 5 (b). The net yield of resonant X-ray emission is calculated to be $(1.5\pm0.5)\times10^{10}$ photons/sr, that corresponds to $(0.428\pm0.08)$ mJ over the full solid angle. Given that the XFEL energy is measured to be $\sim$0.787 mJ, the absorption efficiency in the strong resonant case is approximately $(54 \pm 10)$\%.  In contrast, according to the Henke Tables \cite{Henke1993}, the XFEL absorption in the cold Cu wire with a 10 $\mu$m diameter is 22\%. The increased attenuation of XFEL beam in resonant case is consistent with the opacity calculations shown in figure 4. The measured temporal evolution of the resonant emission suggests that the temperature within the solid-density plasma remains above 600 eV up to 10 ps. 

The history of the resonant emission yield is governed by the dynamics of plasma heating, ionization, and recombination. In conjunction with time-resolved spectroscopic measurements, resonant X-ray absorption imaging can be employed to further probe the spatial distribution of ionization within plasmas, as recently demonstrated in short-pulse laser-driven warm dense matter experiments.\cite{Sawada2023,Sawada2024}. Such experiments significantly advance our understanding of electron transport in solid-density plasmas and holds substantial potential for improving the collisional and ionization models implemented in state-of-the-art PIC and MHD codes.

\subsection{XPCI measuring shock driven compression and expansion}

Lastly, we present the diagnostic results of Fresnel propagation-based X-ray phase-contrast imaging, that is sensitive to the density gradient of sub-$\textmutext$m, aiming to explore the laser driven blast wave in the interaction region and surface return current driven cylindrical compression off the interaction point ranging from tens of ps to hundreds of ps. In contrast to the previous discussed diagnostics of SAXS and RXES with $\sim$10 $\textmutext$m XFEL field of view (FoV) on the 10 $\textmutext$m Cu wire, we switch the X-ray focusing scheme to have $\sim$30 $\textmutext$m FoV illuminating on the target to record the interferences between scattered and unscattered X-rays. The X-ray focal point is calculated to be $\sim$0.76 m in front of the sample with the FWHM focal size $\sim1.7$ $\textmutext$m. The X-ray imaging detector is positioned downstream at a distance of \textit{L} = 6.31 m from the sample plane. This leads to an overall magnification of \textit{M} = (6.31 m + 0.76 m) / 0.76 m = 9.3. The effective Fresnel number with respect to one pixel is then calculated by\cite{Hagemann2021}
\begin{equation}
Fr_{eff}=\frac{\Delta x^2}{\lambda LM} \:, \label{eq:Fresnel}
\end{equation}
with the effecitve pixel size $\Delta x=6.5\mum/7.5=0.867\mum$ and X-ray wavelength $\lambda=0.1512$ nm, giving $Fr_{eff}=10^{-4}$. 

\begin{figure*}
    \centering
    \includegraphics[width=\textwidth]{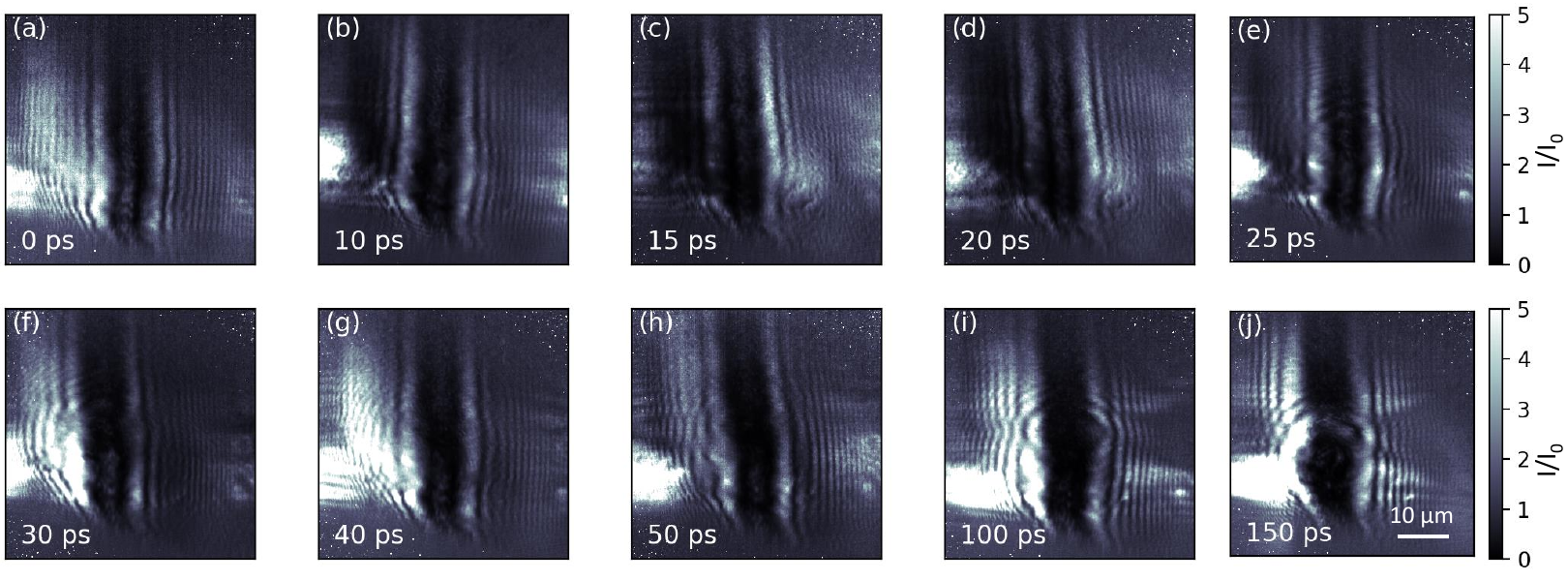}
    \caption{Experimental recorded time resolved XPCI images normalized by the synthetic X-ray flat field from 0 ps (a) to 150 ps (j). Here we switch the XFEL field of view from $\sim$10 $\textmutext$m (used for SAXS and RXES diagnostics) to $\sim$30 $\textmutext$m illuminating on the 10 $\textmutext$m Cu wire by changing the X-ray focusing scheme, enabling the interference between the scattered and free-propagated unscattered X-rays. The scale bar applies to all panels. }
    \label{fig:fig6}
\end{figure*}

\begin{figure*}
    \centering
    \includegraphics[width=\textwidth]{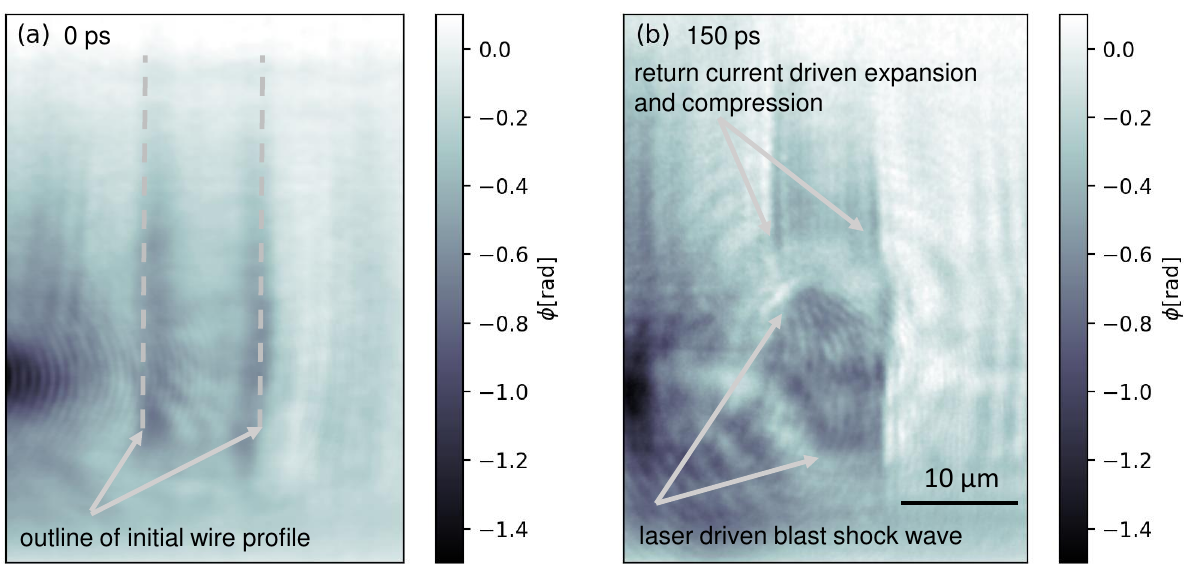}
    \caption{Reconstructed phase maps from XPCI measurements at 0 ps (a) and 150 ps (b). Notice that distances from the focus to the wire and from the focus to the detector are set 0.1 m and 5.367 m respectively, to achieve better reconstruction quality compared to using the nominal propagation parameters.
    }
    \label{fig:fig7}
\end{figure*}

Figure 6 shows the time-resolved XPCI images normalized by a synthetic X-ray flat field (X-ray illumination intensity) from 0 ps to 150 ps. For each delay, the synthetic flat field is computed based on the principal component analysis (PCA) of a series of empty XFEL pulses of post shot when the target is totally destroyed by the ReLaX laser\cite{Hagemann2021}. Since the hot and post shots for each target happen in a short time window ($\sim$two minutes), it is presumed that the principle components of the XFEL illumination field $C_i$ remain unchanged. The PCA algorithm is then in principle to solve a minimization problem. Given a measured intensity image \textit{Y} (hot shot), we calculate the combination of weight factors $w_i$ of the PCA components $C_i$ that minimizes $\lvert Y- \sum_i w_i C_i\rvert$. This minimization can be approximated by the projection of \textit{Y} onto the \textit{C}'s, computed by a scalar product, which yields the $w_i$. With these steps, the synthetic flat field of each hot shot is computed. The whole algorithmic process has been thoroughly detailed in the previous work\cite{Hagemann2021}. 

As shown in figure 6, we can clearly see the Fresnel fringes formed by interference between scattered and free-propagated unscattered X-rays. The fringes exhibit axially asymmetric contrast on either side of the wire due to the XFEL spatial jitter and non-uniform distribution. The bright spots outside the primary beam and the twin fringes are the artifacts caused by the non-ideal beam shape and normalization algorithm. Nevertheless, the XPCI pattern shows significant changes with the time delay. At 0 ps, the interference fringes remain nearly straight, indicating that no significant target deformation has occurred due to hydrodynamic expansion and compression. As time delay increases, the fringes progressively exhibit greater degree of curvature, indicating target deformation due to the hole boring process caused by the laser driven blast wave at the interaction point. Such hole boring process was also observed by the tilt SAXS signal of the scattering surface in our previous work\cite{Kluge2023}. At 150 ps, the XPCI pattern is altered significantly, showing a nearly circular curvature with distinct expanding features, indicating the wire target is partially destroyed. Furthermore, the curved fringes observed outside the primary beam region suggest wire surface ablation which can cause surface compression and expansion induced by the heating of flowing return current\cite{Yang2024}.

In order to precisely retrieve the key plasma parameters such as density and temperature, the ideal approach is to reconstruct the attenuation and phase-shift maps of the XFEL beam crossing the wire in the real space that are encoded to the contrast fringes of the X-ray imaging shown in figure 6. Here, phase retrieval is performed using the efficient iterative algorithm of alternating projections\cite{Hagemann2021}, enabling reconstruction of the phase shift maps shown in figure 7. The phase shift is then expected to be proportional to the projected density along the XFEL beam path\cite{Montgomery2023}. As we can see, the initial profile of 10 $\textmutext$m Cu wire is clearly outlined without deformation by the intense laser pulse at 0 ps yet. At 150 ps, the laser-driven blast shock wave traverses radially through the entire thin wire, breaks out at the rear surface while continues to propagate transversely. The average velocity of the blast shock is estimated faster than 67 km/s, which is in agreement with the observation of direct X-ray imaging on 25 $\textmutext$m Cu wire irradiated by the same ReLaX laser\cite{Garcia2024}. Based on the theoretical prediction $v_{shock}^{thermal} =\sqrt{3P/\rho_0}$ where $\rho_0$ is the initial mass density of Cu\cite{Santos2017}, the thermal pressure of the blast shock is higher than 150 Mbar in the interaction region. Furthermore, the slight expansion of the wire profile away from the interaction point indicates plasma ablation which can be driven by the surface return current heating. This phenomenon is associated with cylindrical compression propagating toward the wire core, as predicted and more clearly observed in our recent direct X-ray imaging results, achieved by placing a focusing compound refractive lens stack between the target and the detector\cite{Yang2024,Garcia2024}. The weak visibility of compression feature in the phase map is attributed to the limited reconstruction quality due to the uncertainty of the propagation distances and artifacts of the flat fielded PCI caused by the beam clipping seen the bright spots in figure 6.

Alternatively, one could simulate the forward synthetic XPCI imaging based on the Fresnel diffraction integral formula\cite{Goodman2017} to best fit the experimental measurements:
\begin{equation}
E(x,y,z)=\frac{1}{i \lambda} \iint_{-\infty}^{\infty} E(x^{\prime},y^{\prime},0) \frac{e^{i k r}}{r} \frac{z}{r}dx^{\prime} dy^{\prime}, \label{eq:Diffraction}
\end{equation}
where $r=\sqrt{(x-x^{\prime})^2 + (y-y^{\prime})^2+z^2}$, $i$ is the imaginary unit, $k$ is the wavenumber and $E(x^{\prime},y^{\prime},0)$ is the electric field of X-ray just exits the target plane. Such forward imaging simulations require inputs of density, opacity and heating depth, which are typically predicted through a combination of PIC and MHD, or hybrid simulations, and are therefore highly model-dependent. This will be investigated in a follow-up study, as it is beyond the scope of the present demonstration work. 

Besides investigating shock and pinch physics, we also aimed to probe density modulations arising from the seeding and growth of kinetic and hydrodynamic instabilities using propagation-based XPCI. However, such phenomena have not been observed here, likely due to limitations in imaging resolution. Employing the seeding mode of the XFEL with a narrower bandwidth and nanofocused beam with a higher geometric magnification could improve the imaging resolution and quality for such scientific purposes\cite{Schropp2015}. It is also noticed that although the direct X-ray imaging setup\cite{Garcia2024} provides intuitive and straightforward observations of laser-driven blast wave and transient surface return current-induced compression, offering a clear advantage over the phase retrieval results from XPCI in this setup, its phase reconstruction could suffer from phase errors caused by the fabrication imperfections in the imaging CRL stack placed between the sample and detector\cite{Celestre2020}. Additionally, direct X-ray imaging is limited in its application to low-density samples due to constraints of absorption contrast. These limitations could be addressed by the newly commissioned Talbot X-ray imaging technique at the XFEL facility, which incorporates an additional Talbot grating into the direct X-ray imaging setup\cite{Galtier2025}.

\section{Conclusions}

In this study, we successfully demonstrate the capability to capture multi-scale spatio-temporal dynamics of electron transport, ranging from a few nm to sub-$\textmutext$m and from sub-ps to hundreds of ps, in a hot solid density plasma created by an optical relativistic 3J/30fs/100TW laser. This is achieved using a high-brightness XFEL within the novel pump-probe platform at the European XFEL-HED station. By simultaneously employing multiple X-ray diagnostics of SAXS, XPCI and RXES, we are able to probe the preplasma expansion driven by the leading edge of relativistic laser pulse, ultra-fast heating and ionization dynamics at the arrival of the main pulse, and hydrodynamic evolution to 150 ps after the laser wire interactions. The decreasing tread of edge streak SAXS intensity sensitive to a few nm structure changes, reveals that the preplasma is formed and expanded at least 30 ps (corresponding to the intensity of $\sim5\times10^{13}$ W/cm$^{2}$) prior to the arrival of peak pulse. Furthermore, we present time-resolved RXES measurements, achieved by resonantly XFEL pumping specific charge state in ReLaX laser-driven hot dense Cu plasmas. The intensity of the resonant X-ray emission directly reflects the population of the corresponding ionic charge states Cu$^{22+}$, thereby providing valuable insights into the temporal evolution of heating, ionization, and recombination dynamics. Finally, we present the temporal evolution of XPCI  measurements and corresponding phase reconstruction results, which reveal the hydrodynamic expansion and compression induced by shock waves originating from direct laser ablation and transient surface return currents sensitive to the sub-$\mum$ scale density gradients. Further improvements to the experimental capability, such as accessing the lower SAXS regime, implementing resonant absorption imaging, and utilizing Talbot X-ray imaging are also proposed.

To the best of our knowledge, this work constitutes the first comprehensive investigation of complex solid density plasma dynamics throughout the entire process of relativistic laser–solid interactions within a single experiment, enabled by the unprecedented probing capabilities of the XFEL. These findings offer a valuable experimental benchmark for PIC and MHD simulations, which are widely employed in the high energy density physics community. The results are also benefit to the advanced applications of laser plasma-based particle accelerator, the creation of high energy density matter, and could provide useful insights to understand the planetary science, astrophysics and fusion energy research.

\section*{Acknowledgement}

We acknowledge European XFEL in Schenefeld, Germany, for provision of X-ray free-electron laser beamtime at HED SASE2 under proposal number 3129 and would like to thank the staff for their assistance. The authors are indebted to the HIBEF user consortium for the provision of instrumentation and staff that enabled this experiment. T. Engler acknowledges the funding of Grant-No. HIDSS-0002 DASHH (Data Science in Hamburg - Helmholtz Graduate School for the Structure of Matter). The data recorded for the experiment at the European XFEL are available upon reasonable request\cite{Data3129}. 

\section*{Authors contributions}

L.H. lead the experiment. L.H., M.S, A.L.G., T.K., M.M., T.P., H-P.S., T.T., U.Z. and T.E.C conceived the experimental setup. L.H., M.S., L.Y., T.E., X.P., C.B., E.B., A.L.G., S.G., M.H., H.H., M.K., W.L., M.M., M.N., M.O., \"{O}.\"{O}., A.P., T.P., L.R., M.R., H-P.S., J-P.S., M.T., T.T. and T.E.C performed the experiment. L.H., M.S., L.Y., O.H., J.H., T.E., X.P., Y.C., T.K., R.A., E.E., A.L.G., P.H., T.P., S.S., E.T., T.T. U.Z., and T.E.C. analyzed the data. L.H. wrote the original manuscript draft. All authors discussed the results and revised the manuscript. 

\bibliography{main}

\end{document}